\documentclass{Interspeech2024}

\interspeechcameraready

\usepackage{cite}
\usepackage{subcaption}
\usepackage{multirow}
\usepackage{xcolor}
\usepackage{comment}
\usepackage{xurl}
\def\x{{\mathbf x}}
\def\y{{\mathbf y}}
\def\h{{\mathbf h}}

\def\A{{\mathbf A}}
\def\B{{\mathbf B}}
\def\C{{\mathbf C}}

\def\W{{\mathbf W}}

\def\R{{\mathbb R}}

\title{Exploring the Capability of Mamba in Speech Applications}

\name[affiliation={1}]{Koichi}{Miyazaki}
\name[affiliation={1,2}]{Yoshiki}{Masuyama}
\name[affiliation={1}]{Masato}{Murata}

\address{
  $^1$CyberAgent, Inc., Japan\\
  $^2$Tokyo Metropolitan University, Japan
}
\email{miyazaki\_koichi\_xa@cyberagent.co.jp}


\begin{document}
\maketitle

\begin{abstract}
This paper explores the capability of Mamba, a recently proposed architecture based on state space models (SSMs), as a competitive alternative to Transformer-based models.
In the speech domain, well-designed Transformer-based models, such as the Conformer and E-Branchformer, have become the de facto standards.
Extensive evaluations have demonstrated the effectiveness of these Transformer-based models across a wide range of speech tasks.
In contrast, the evaluation of SSMs has been limited to a few tasks, such as automatic speech recognition (ASR) and speech synthesis.
In this paper, we compared Mamba with state-of-the-art Transformer variants for various speech applications, including ASR, text-to-speech, spoken language understanding, and speech summarization.
Experimental evaluations revealed that Mamba achieves comparable or better performance than Transformer-based models, and demonstrated its efficiency in long-form speech processing.

\end{abstract}

\section{Introduction}
Speech processing has witnessed significant performance improvements with the recent progress in end-to-end sequence-to-sequence models~\cite{e2easr,e2est,seamless}.
The key to this advancement is Transformer~\cite{transformer} with self-attention that captures the global context and allows parallel training.
Consequently, the performance of various speech processing tasks has been improved over long short-term memory-based models~\cite{transformer-vs-rnn}.

Building upon the success of Transformers, several variants have been tailored for speech processing tasks~\cite{conformer,ebranchformer,zipformer}.
Conformer~\cite{conformer} and E-Branchformer~\cite{ebranchformer} have become de facto standards in this field.
Conformer combines a self-attention mechanism with convolutional neural networks to capture both global and local contextual information in a cascaded manner.
This hybrid architecture demonstrated superior performance in automatic speech recognition (ASR) and text-to-speech (TTS).
Meanwhile, E-Branchformer adopts a parallel structure to extract global and local contexts, and subsequently merges the outputs of the two branches.
E-Branchformer has achieved state-of-the-art results on various benchmarks~\cite{conformer-vs-e-brachformer}.

Despite the success of Transformers, they suffer from the quadratic time and memory complexity in the vanilla attention mechanism.
This prevents models from scalability to long-form speech and motivates exploring alternative to Transformers~\cite{branchformer,summarymixing}.
Towards efficient modeling of long-range dependencies, neural state space models (SSMs) have emerged as a promising alternative architecture.
In particular, SSMs leverage a state to represent the past sequences instead of attending the entire sequences at each time step.
This procedure can be performed in parallel with a sub-quadratic complexity by using tailored algorithms.
SSMs have exhibited promising results in several speech processing tasks, such as ASR~\cite{dssformer,s4decoder,mssm,s4former}, speech synthesis~\cite{sashimi}, and speech enhancement~\cite{s4m,s4ndunet}.
Most SSMs, including S4~\cite{s4}, have mimicked a time-invariant system, which inherently limiting their ability for input-dependent processing represented by selective copying and induction heads~\cite{mamba}.

To overcome this limitation, Mamba introduced a selection mechanism that parameterizes the SSM parameters based on the input-dependent processing~\cite{mamba}.
This modification has yielded promising results in various fields such as natural language processing (NLP)~\cite{mambabyte}, and computer vision (CV)~\cite{vision-mamba}.
Although Mamba seems to be competitive alternative to Transformers, its superiority in general speech applications was not comprehensively validated in the original paper~\cite{mamba}.

In this paper, we investigate the efficacy of Mamba on various end-to-end speech processing applications: ASR, TTS, spoken language understanding (SLU), and speech summarization (SUMM).
Our experimental evaluation covers various types of tasks and long-form speech processing.
Experimental results show that Mamba achieves comparable performance to Conformer/E-Branchformer on various datasets, even outperforming Transformers on long-form speech processing.
We also stress that Mamba is directly applicable to long-form SUMM due to its sub-quadratic complexity, even when Conformer faces the issue of out-of-memory.

\begin{figure*}[t]
    \centering
    \includegraphics[width=0.95\linewidth]{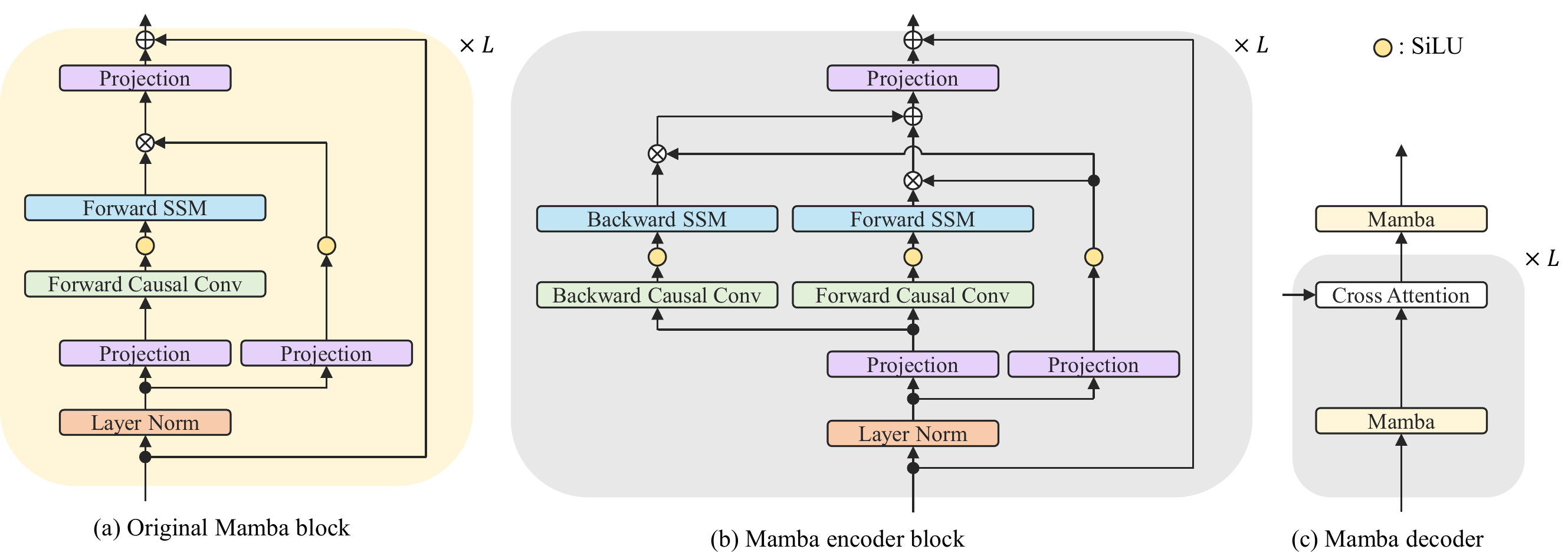}
    \vskip -0.1in
    \caption{Architecture of the Mamba block. (a) The original Mamba block. (b) The Mamba encoder block extends the original Mamba block in a bidirectional design. This modification allows for capturing past and future contexts in the input sequences. (c) The Mamba decoder. To bridge the encoder output, we employed a cross-attention after an original Mamba block.}
    \label{fig:mamba}
    \vskip -0.2in
\end{figure*}
\section{Mamba}

In this section, we briefly review Mamba, which is a recently proposed SSM, as a promising sequence-to-sequence model.
SSMs are inspired by well-established frameworks such as Kalman filters and hidden Markov models~\cite{s4,s4d,s5}.

An SSM maps an input sequence $\x \in \R^{D}$ to $\y \in \R^{D}$ in an element-wise manner.
In detail, the discrete SSM with element-wise latent state $\h_d \in \R^{N}$ can be formulated as follows\footnote{Note that ~\cite{mamba} explains zero-order hold discretization; however, we used an approximated version followed by the official implementation.}:
\begin{align}
    \h_{t, d} &= \Bar{\A} \h_{t-1, d} + \Bar{\B} x_{t, d}, \\
    y_{t, d} &= \C \h_{t, d}, \\
    \Bar{\A}, \Bar{\B} &= \exp (\Delta \A), \Delta \B,
\end{align}
where $\A \in \R^{N\times N}$, $\B \in \R^{N\times 1}$, $\C \in \R^{1 \times N}$, and $\Delta \in \R_+$ represent continuous SSM parameters~\cite{s4}.
We assume that $\A$ is diagonal for simplicity and computational efficiency.
One of the key contributions of Mamba is introducing a selection mechanism that allows SSM to be input-dependent.
That is, rather than directly optimizing the SSM parameters, we optimize additional parameters ($\W_B$, $\W_C$, and $\W_\Delta$) that compute the SSM parameters as follows:
\begin{align}
    \B_t, \C_t &= \W_B \x_t, (\W_C \x_t)^{\mathsf{T}}, \\
    \Delta_t &= \text{softplus}(\W_{\Delta} \x_t),
\end{align}
where $\W_B \in \R^{N\times D}$, $\W_C \in \R^{N\times D}$, $\W_\Delta \in \R^{1\times D}$, softplus function refers to $\log(1+\exp(x))$, and $(\cdot)^{\mathsf{T}}$ denotes the transpose.
By making the SSM input-dependent, however, the efficient algorithms that utilize global convolution~\cite{s4} are no longer applicable.
To address this limitation, Mamba uses parallel scan~\cite{blelloch-scan,s5} and hardware-efficient algorithms.

In this study, we used Mamba in the encoder and decoder, where we modified the original Mamba block as illustrated in Figure~\ref{fig:mamba}. Each block had a layer normalization~\cite{layernorm} and residual connection.
In the inner block, the input signal mapped $\R^{D_\text{in}} \mapsto \R^{E D_\text{in}}$ through an input projection layer, then applied causal convolution and SSM layer. 
Thereafter, the processed signal mapped back $\R^{E D_\text{in}} \mapsto \R^{D_\text{in}}$ through an output projection layer.
As the SSM has a causal operation, we extended bidirectional design similar, to that ofbidirectional RNNs~\cite{brnn}.
To bridge the encoder output to the decoder, we employ cross-attention after each Mamba block.
Note that SSM handles positional information implicitly, we removed the positional encoding used in the Transformer-based encoder and decoder.
We employed the official Mamba codebase\footnote{\url{https://github.com/state-spaces/mamba}}.
We used the S4D-Real~\cite{s4d} initialization such that $\A_n = -(n+1)$, state size $N$ set to 16, expansion factor $E$ set to 4, and initial $\Delta$ parameters are uniformly sampled from $[0.001, 0.1]$ for all experiments.

\begingroup
\setlength{\tabcolsep}{2pt}

\begin{table}[t]
  \caption{WER (\%) for different encoder and decoder architectures on LibriSpeech 100h test sets.
  CTC and AED are performed with greedy search and beam search, respectively. Real-time factor (RTF) is calculated using a single A100 GPU. All results are obtained without an external language model.}
  \label{tab:ctc}
  \vskip -0.1in
  \centering
  \resizebox {\linewidth} {!} {
  \begin{tabular}{cc|c|ccc}
    \toprule
    \multicolumn{2}{c|}{\textbf{Model}} & & \multicolumn{3}{c}{Results$\downarrow$} \\
    Encoder & Decoder & Params & clean & other & RTF \\
    \midrule
    \midrule
    \textbf{CTC} \\
    \midrule
        \quad Transformer & N/A & 17.3 & 12.8 & 28.1 & 0.118 \\
        \quad Conformer & N/A & 27.0 & 9.8 & 23.3 & 0.193 \\
        \quad E-Branchformer & N/A & 26.4 & 9.5 & \textbf{22.9} & 0.189 \\
        \quad (uni-)Mamba & N/A & 24.2 & 15.7 & 33.6 & \textbf{0.117} \\
        \quad (bi-)Mamba & N/A & 26.3 & \textbf{9.1} & 23.5 & 0.152 \\

    \midrule
    \textbf{AED} \\
    \midrule
        \quad E-Branchformer & Transformer & 38.5 & 6.4 & 17.0 & 0.453 \\
        \quad E-Branchformer & S4 & 34.9 & 6.3 & 16.5 & 0.360 \\
        \quad E-Branchformer & (uni-)Mamba & 36.7 & \textbf{6.1} & \textbf{16.5} & 0.357 \\ 
        \quad (bi-)Mamba & Transformer & 38.3 & 6.6 & 18.9 & 0.351 \\
        \quad (bi-)Mamba & S4 & 34.8 & 6.5 & 18.6 & 0.349 \\
        \quad (bi-)Mamba & (uni-)Mamba & 36.6 & 6.5 & 18.5 & \textbf{0.346} \\ 
    \bottomrule
  \end{tabular}
  }
  \vskip -0.2in
\end{table}
\endgroup

\begingroup
\setlength{\tabcolsep}{2pt}

\begin{table*}[t]
  \caption{ASR results on various datasets using hybrid CTC/Attention model. Token refers to the input and output token type. Char and BPE represent character and byte pair encoding, respectively. Params refers to the total number of parameters ($\times 10^6)$. $\dagger$ means the result with a shallow fusion of a Transformer language model.
  $\ddagger$ means Conformer encoder is used instead of E-Branchformer as the training was failed.
  Note that E-Branchformer-Transformer is reproduced by the provided recipe in ESPnet2.
  }
  \label{tab:main-results}
  \vskip -0.1in
  \centering
  \resizebox {0.97\linewidth} {!} {
  \begin{tabular}{c @{\hskip 2pt} c @{\hskip 2pt} c|c|c|c|c|c}
  \toprule
  \multirow{2}{*}{Dataset} & \multirow{2}{*}{Token} & \multicolumn{1}{c}{\multirow{2}{*}{Metric}} & \multicolumn{1}{c}{\multirow{2}{*}{Evaluation Sets}} & \multicolumn{2}{c}{E-Branchformer-Transformer} & \multicolumn{2}{c}{E-Branchformer-Mamba}\\
  \cmidrule(lr){5-6}
  \cmidrule(lr){7-8}
  & & \multicolumn{1}{c}{} & \multicolumn{1}{c}{} & \multicolumn{1}{c}{Params} & \multicolumn{1}{c}{Results $\downarrow$} & \multicolumn{1}{c}{Params} & \multicolumn{1}{c}{Results $\downarrow$} \\
  \midrule
  \midrule
  AISHELL~\cite{aishell-corpus} & Char & CER & dev / test & 45.7  & \textbf{4.2} / \textbf{4.4} & 43.9 & \textbf{4.2} / 4.6 \\
  GigaSpeech~\cite{gigaspeech} & BPE & WER & dev / test & 148.9 & \textbf{10.5} / \textbf{10.6} & 153.7 & 10.7 / 10.7 \\
  CSJ~\cite{csj} & Char & CER & eval1 / eval2 / eval3 & 146.3 & \textbf{3.5} / \textbf{2.7} / \textbf{2.9} & 151.1 & 3.7 / 2.8 / \textbf{2.9} \\
  LibriSpeech 100h~\cite{librispeech-corpus} & BPE & WER & \{dev,test\}\_\{clean,other\} & 38.5 & 6.3 / 17.0 / 6.4 / 17.0 & 36.7 & \textbf{6.0} / \textbf{16.2} / \textbf{6.1} / \textbf{16.5} \\
  LibriSpeech 960h~\cite{librispeech-corpus} & BPE & WER & \{dev,test\}\_\{clean,other\} & 148.9 & $\dagger$ \textbf{1.7} / \textbf{3.6} / 1.9 / \textbf{3.9} & 153.7  & $\dagger$ 1.7 / \textbf{3.6} / \textbf{1.8} / \textbf{3.9} \\
  TEDLIUM2~\cite{tedlium2} & BPE & WER & dev / test & 35.0 & $\ddagger$ 9.1 / 7.4 & 33.3 & $\ddagger$ \textbf{8.1} / \textbf{7.3} \\
  VoxForge~\cite{voxforge} & Char & CER & dt\_it / et\_it & 34.7 & 8.6 / 8.1 & 32.9 & \textbf{8.5} / \textbf{8.0} \\
  \bottomrule
  \end{tabular}
  }
  \vskip -0.2in
\end{table*}
\endgroup

\section{Automatic Speech Recognition}
\subsection{Setups} 

\noindent\textbf{Data.} We used seven diverse ASR datasets that covered various languages, speaking styles, and a range of dataset sizes. The evaluation metrics and dataset split follow the ESPnet recipes\footnote{\url{https://github.com/espnet/espnet}}.

\noindent\textbf{Models.} We compared the attention encoder-decoder (AED) model with different combinations of encoder (E-Branchformer or Mamba) - decoder (Transformer, S4, or Mamba) architectures. 
We explored these combinations to gain insights into the effectiveness of Mamba as an encoder and/or decoder for ASR tasks. For a fair comparison, we set the expansion factor to four in both the Mamba encoder and decoder, resulting in a similar model sizes.  For the small models, the E-Branchformer encoder had 12 blocks, whereas the Mamba encoder had 24 blocks. For the large models, the E-Branchformer encoder had 17 blocks, whereas the Mamba encoder has 30 blocks.
Transformer and Mamba decoder had six blocks in all experiments.

\noindent\textbf{Training.} We followed the ESPnet recipes for data preparation, training, decoding, and evaluation. The data augmentation and hyper-parameters were the same as those provided in the recipe for the E-Branchformer.
We observed that the Mamba model was sometimes unstable during training and tended to overfit. We thus added dropout regularization after the input and output projection layers.
To stabilize training, we used AdamW optimizer~\cite{adamw} instead of Adam~\cite{adam} and set a stricter dropout probability of 0.2.
All models were trained on A100 GPUs.

\subsection{Results}
Table~\ref{tab:ctc} lists the ASR results for various combinations of encoders and decoders on the LibriSpeech 100h dataset. We compared the encoder-only models using CTC.
The CTC results reveal that making Mamba bi-directional leads significant performance improvements. However, it did not surpass the Conformer/E-Branchformer in test\_other. This suggests the Mamba model may be less robust in encoding acoustic features.
The AED results demonstrate that using an SSM-based decoder yields a better performance compared to the Transformer decoder.
Table~\ref{tab:main-results} lists ASR results across various benchmarks. We compared the E-Branchformer-Transformer with the E-Branchformer-Mamba, which performs the best, as summerized in Table~\ref{tab:ctc}.  
E-Branchformer-Mamba performed comparably to E-Branchformer-Transformer in various scenarios.

\subsection{Discussion}
\begin{figure}[t]
    \centering
    \vskip -0.1in
    \includegraphics[width=0.98\linewidth]{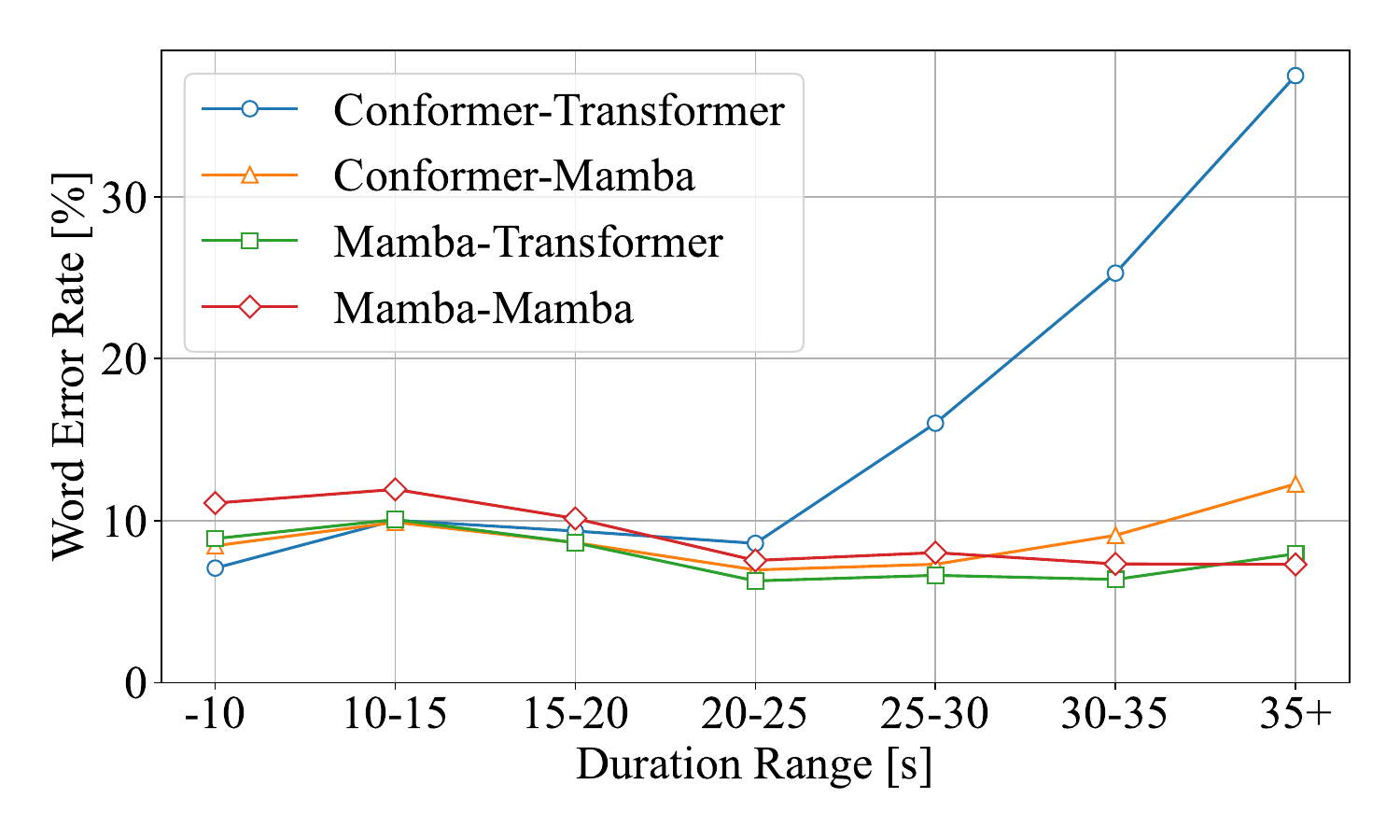}
    \vskip -0.2in
    \caption{Long-form ASR results on TEDLIUM2.}
    \label{fig:long-form_asr_on_tedlium2}
    \vskip -0.2in
\end{figure}

Figure~\ref{fig:long-form_asr_on_tedlium2} shows the performance of long-form speech recognition.
The models were trained using the TEDLIUM2 dataset and evaluated using three concatenated consecutive speech segments.
The Conformer-Transformer exhibited a gradual degradation in recognition performance as the length of the evaluated speech exceeded 25 s.
We successfully mitigate this performance degradation by incorporating Mamba into either the encoder or decoder,
This observation is consistent with recent findings that highlighting the challenges associated with positional embedding in processing long sequence lengths.
Therefore, it supports the efficacy of the models employing SSMs for enhanced long-form speech recognition.

\section{Text-to-Speech}
To investigate the capability of Mamba for generative tasks, we performed TTS experiments using LJSpeech~\cite{ljspeech} dataset.

\subsection{Setups}

\noindent\textbf{Data.}
LJSpeech dataset contains 24 hours of audiobook speech uttered by a single female speaker. The dataset had 13,100 utterances, and we split the dataset into 12600/250/250 utterances each for training, development, and evaluation set, respectively.

\noindent\textbf{Models.}
We used Conformer-FastSpeech2~\cite{espnet2-tts} as the baseline, which is an extension of the FastSpeech2~\cite{fastspeech2} model using Conformer blocks instead of Transformer blocks.
To assess the  effectiveness of Mamba in TTS, we integrated Mamba blocks into the FastSpeech2 backbone (Mamba-FastSpeech2).
Since FastSpeech2 is a non-autoregressive model, we used a bidirectional Mamba for both the encoder and decoder modules.
To synthesize waveform from the generated acoustic features, we use a HiFi-GAN vocoder~\cite{hifi-gan}. The HiFi-GAN model is trained using the same dataset split as the baseline.

\noindent\textbf{Training.}
We followed the ESPnet2 recipe but reduced the total training steps from $1.0 \times 10^6$ to $1.0 \times 10^5$.
This reduction is sufficient to achieve reasonable quality while significantly reducing the computational cost and training time.
We used g2p\_en\footnote{\url{https://github.com/Kyubyong/g2p}} as a grapheme-to-phoneme function.
Conformer-FastSpeech2 has four Conformer blocks in the encoder and decoder modules.
Mamba-FastSpeech2 has eight Mamba encoder blocks in the encoder and decoder modules.
We used the duration labels extracted using the Montreal Forced Aligner toolkit~\cite{mfa}.

\subsection{Results}
\begin{figure}[t]
    \centering
    \vskip -0.1in
    \includegraphics[width=0.98\linewidth]{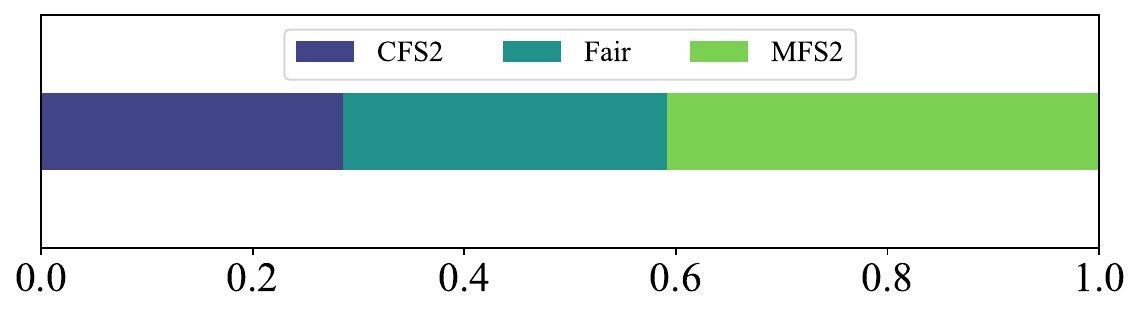}
    \vskip -0.2in
    \caption{Preference test results on CFS2 vs. MFS2.}
    \label{fig:preference_test}
    \vskip -0.2in
\end{figure}

We conducted subjective and objective evaluations to assess speech quality. 
For the subjective evaluation, we performed a 5-scale mean opinion score (MOS) test and a preference test via Amazon Mechanical Turk with 50 participants.
They evaluated randomly selected 50 audio samples from each method and five random audio pairs with the same utterance. 
For the objective evaluation, we followed the evaluation process in ESPnet-TTS, including the mel-cepstral distortion (MCD), log-$F_o$ root-mean-square error (log$F_o$), and CER.
Table~\ref{tab:tts_results} and Figure~\ref{fig:preference_test} show the TTS results, indicating that Mamba-FastSpeech2 has comparable performance on both subjective and objective evaluations.

\begin{table}[t]
    \caption{\it{TTS results. GT (mel) refers to a reconstructed sample using ground truth mel-spectrogram with HiFi-GAN vocoder. CFS2 and MFS2 denote the synthesized speech samples from Conformer-FastSpeech2 and Mamba-FastSpeech2, respectively. CI represents the 95 \% confidence interval}}
    \vskip -0.1in
    \centering
    \vspace{2mm}
    \scalebox{0.80}[0.80]{
    \begin{tabular}{l|cccc}
        \toprule
        Method          & MCD [dB] $\downarrow$ & log$F_o$ $\downarrow$ & CER (\%)$\downarrow$ & MOS $\uparrow$ $\pm$CI \\
        \midrule
        \midrule
        GT(mel)         &  3.75 & 0.156 & 1.1 & 4.06 $\pm$ 0.11      \\
        \midrule
        CFS2            & \textbf{6.51} & \textbf{0.217} & \textbf{1.7} & 3.72 $\pm$ 0.12 \\
        MFS2            & 6.54 & 0.219 & 1.9 & \textbf{3.76 $\pm$ 0.12} \\
        \bottomrule
    \end{tabular}
    }
    \label{tab:tts_results}
\end{table}

\section{Spoken language understanding}

To demonstrate the effectiveness of Mamba in predicting high-level semantics, we performed SLU experiments on SLURP~\cite{slurp-corpus} and SLUE~\cite{shon2022slue} following ESPnet-SLU~\cite{espnet-slu}.

\subsection{Setups}
\noindent\textbf{Data.}
SLURP~\cite{slurp-corpus} contains utterances of single-turn user interactions with a home assistant, where each recording is annotated with a scenario, action, and entities.
We performed intent classification and entity prediction, where the intent corresponds to the scenario and action pair.
SLUE~\cite{shon2022slue} is a low-resource benchmark for sentiment analysis and named entity recognition.
As the annotations for the evaluation set are not public, we report the results for the development set.

\noindent\textbf{Models.}
AED models were used to jointly predict the label for SLU and the corresponding transcription~\cite{espnet-slu}.
We investigated the performance using different encoder architectures: Conformer, E-branchformer, and Mamba.
We used the popular log-mel filter-bank as the front-end.
The baseline model followed the configurations adopted in \cite{conformer-vs-e-brachformer} and \cite{pengIntegrationSLU} for SLURP and SLUE, respectively.
The Mamba encoder followed a small model in the ASR experiment and aligned the number of parameters to those of the baseline models.
The Transformer decoder consisted of six blocks.

\noindent\textbf{Training.}
We followed the ESPnet recipes for data preparation, training, and inference.
For SLURP, we set the maximum learning rate to $2.0 \times 10^{-4}$ with Adam optimizer following the baseline configuration.
For SLUE, we used AdamW optimizer and adjusted the learning rate to $2.0 \times 10^{-3}$ and $5.0 \times 10^{-4}$ for sentiment analysis and named entity recognition, respectively.

\begin{table}[t]
  \caption{SLU results with different encoders.
  SLURP shows intent accuracy (\%) and SLU-F1 (\%).
  SLUE shows macro F1 (\%) for sentiment analysis and named entity recognition.
  }
  \label{tab:slu}
  \vskip -0.1in
  \centering
  {\footnotesize
  \begin{tabular}{c|ccc}
    \toprule
    Dataset & Conformer & E-branchformer & Mamba \\
    \midrule
    \midrule
    SLURP & 86.1 / 77.4 & 86.8 / 78.0 & \bf{88.1} / \bf{78.3} \\
    SLUE & 34.2 / 39.4 & 33.5 / 40.5 & \bf{35.0} / \bf{41.9} \\
    \bottomrule
  \end{tabular}
  }
\end{table}

\subsection{Results}
Table~\ref{tab:slu} compares the SLU results for the different encoder architectures.
Mamba consistently performed better than Transformers with a similar model sizes.
In particular, on SLURP, we computed the confidence intervals for intent accuracy using the official Python toolkit.
We confirmed that the confidence interval for Mamba was $(87.5, 88.7)$, while that for E-Branchfromer was $(86.2, 87.4)$.
This result indicates that Mamba has the potential to outperform Transformers in semantic tasks.

\section{Speech summarization}
End-to-end abstractive speech summarization aims to generate a short summary from a long-form speech, which requires modeling the long-range dependencies~\cite{Roshan2022,Kano2023}.
Since we need to handle long-form speech in the encoder, the quadratic complexity of the attention mechanism is problematic. Hence, we explore the benefit of Mamba in the encoder.

\subsection{Setups}
\noindent\textbf{Data.}
We used the How2 corpus~\cite{how2} containing 2000 hours YouTube videos and their descriptions.
Utterance-level and entire video-level sub-sets were provided for ASR and SUMM, respectively. 
As described in \cite{espnet-summ}, we used 40-dimensional filter-bank and 3-dimensional pitch features were used as inputs.

\noindent\textbf{Models.}
We replaced the Conformer encoder in the baseline AED with a 24-block Mamba encoder, as its attention mechanism is the main bottleneck in computational complexity.
The number of parameters was aligned with the baseline models. 
The Transformer decoder comprised six blocks for both models.
Our entire model had $96.4 \times 10^6$ parameters, whereas the baseline Conformer model had $97.7 \times 10^6$ parameters.
We also investigated the performance by replacing the Transformer decoder with a Mamba decoder.

\noindent\textbf{Training.}
We pre-trained AED on utterance-level ASR before fine-tuning on SUMM over entire video~\cite{espnet-summ}.
We pre-trained the Mamba encoder with the AdamW optimizer, where the peak learning rate was $2.0 \times 10^{-3}$.
The AED was subsequently fine-tuned on SUMM with an initial learning rate of $1.0 \times 10^{-4}$.
During the fine-tuning of the Conformer model, we trimmed input audio at $100$ s  owing to the out-of-memory issue on A100~\cite{Roshan2022,espnet-summ}.
In contrast, our Mamba model can leverage $600$ s audio thanks to its sub-quadratic complexity.

\subsection{Results}
We evaluated the generated summarization by ROUGE-L (R-L)~\cite{rouge}, METOR (MTR)~\cite{meteor}, and BERTScote (BSc)~\cite{bertscore}, as proxies for human evaluation.
Table~\ref{tab:summ} lists the performances of different architectures and input lengths.
The proposed Mamba model outperformed the Conformer model even with the same input length, and its performance was further improved by leveraging a longer input.
While the Mamba model resulted in slightly worse performance compared to another sub-quadratic encoder known as FNet~\cite{fnet}, this result indicates the potential of Mamba in long-form speech processing tasks.

\begin{table}[t]
  \caption{SUMM results on How2 dataset.}
  \label{tab:summ}
  \vskip -0.1in
  \centering
  {\footnotesize
  \begin{tabular}{c|ccc}
    \toprule
    Method & R-L $\uparrow$ & MTR $\uparrow$ & BSc $\uparrow$ \\
    \midrule
    \midrule
    Conformer-Transformer (100s) & 60.5 & 32.2 & 92.5 \\
    Mamba-Transformer (100s) & 62.3 & 33.5 & 92.9 \\
    Mamba-Transformer (600s) & 62.9 & 33.8 & 93.1 \\
    Mamba-Mamba (600s) & 62.7 & \bf{33.9} & 93.0 \\
    \midrule
    Conformer-Transformer (100s)~\cite{espnet-summ} & 62.3 & 30.4 & 93.0 \\
    FNet-Transformer~\cite{espnet-summ} & \bf{64.0} & 32.7 & \bf{93.5} \\
    \bottomrule
  \end{tabular}
  }
  \vskip -0.1in
\end{table}

\section{Conclusion}
In this paper, we explored the capability of Mamba in various speech applications, including ASR, TTS, SLU, and SUMM.
The experimental results demonstrated that Mamba achieved a performance comparable to that of state-of-the-art Transformer variants across a wide range of benchmarks.
In particular, Mamba exhibited advantages in long-form ASR and SUMM, not only in recognition performance but also in robustness and memory efficiency.
We plan to conduct a detailed analysis focusing on the differences in behavior between Mamba and Transformer variants in future work.

\clearpage

\bibliographystyle{IEEEtran}
\bibliography{mybib}

\end{document}